\documentclass[prb,twocolumn,showpacs]{revtex4}

\usepackage{epsfig}
\begin{document}


\title{Kink-Antikink Unbinding Transition in the Two Dimensional 
Fully Frustrated XY Model}

\author{Peter Olsson$^1$ and S. Teitel$^2$ }
\affiliation{$^1$Department of Theoretical Physics, Ume{\aa} University, 
901 87 Ume\aa, Sweden}
\address{$^2$Department of Physics and Astronomy, University of Rochester, 
Rochester, New York 14627}
\date{\today}

\begin{abstract}
We carry out the first numerical simulations to directly confirm the
existence of a kink-antikink unbinding transition along Ising-like
domain walls in the two dimensional fully frustrated XY model.  
We comment on the possible implications of kink-antikink 
unbinding for the bulk phase transition of the model.
\end{abstract}


\maketitle

\section{Introduction}

The two dimensional (2D) fully frustrated XY model (FFXY) 
\cite{Villain,TJ} is one of the most intriguing of ``simple'' 
statistical mechanics models.
The doubly degenerate checkerboard pattern of vortices
in the ground state leads to an Ising-like discrete Z(2) 
symmetry in addition to the Kosterliz-Thouless-like continuous O(2) 
symmetry associated with the uniform
rotation of all phase angles \cite{KT}.  
It remains controversial whether there are 
two distinct phase transitions $T_{KT}<T_I$, with $T_{KT}$
marking the breaking of the O(2) symmetry and $T_I$ marking the 
breaking of the Z(2) symmetry \cite{double,Olsson1,Olsson2}, or 
rather a single transition 
in which both symmetries are broken simultaneously \cite{single}.  

Recently, Korshunov \cite{Korshunov} presented an argument for a
new interfacial transition $T_w$ in the 2D FFXY model, lying well below the
bulk transition(s), arising from
the unbinding of step excitations of unit height (``kink-antikink pairs'') 
on the domain walls associated with the Z(2) symmetry.  
Korshunov argued that the kink-antikink unbinding transition
leads to a decoupling of phase coherence across domain 
boundaries, supporting the identification of the FFXY with the
coupled XY-Ising model \cite{XYIsing}.  Korshunov further argued that
this effect necessarily leads to the scenario of two distinct bulk transitions,
$T_{KT}<T_I$.

Earlier, Lee and co-workers, first in simulations of the 2D FFXY model 
with Langevin dynamics \cite{LeeXY}, and then in its dual Coulomb gas 
(CG) model with Monte Carlo (MC) dynamics \cite{LeeCG}, found evidence for a 
transition in domain wall 
morphology in simulations of the ordering kinetics of domain growth
following a sudden quench.  They interpreted this as a 
finite temperature roughening transition of the Ising-like domain 
walls.  
Jeon {\it et al.} \cite{Jeon} made similar conclusions in 
simulations of the 2D FFXY with resistively shunted junction dynamics.
Korshunov \cite{Korshunov}, however, has argued that domain walls should be 
rough at all temperatures.

In this paper we present the first direct numerical evidence 
demonstrating the existence of the kink-antikink unbinding transition
at a temperature $T_{w}$ below the bulk transition(s).
In agreement with Korshunov's predictions, we show that phase angles on
opposite sides of the domain wall decouple above $T_{w}$.  
The numerical value we find for $T_{w}$ is comparable to that of the
morphological transition found by Lee {\it et al}. \cite{LeeCG}, 
however we
explicitly demonstrate that domain walls are rough at temperatures
well below $T_{w}$.  This indicates
that the transition seen by Lee and co-workers is really
kink-antikink unbinding, rather than roughening.

\section{The Model}
\subsection{The fully frustrated XY model}

We study the 2D FFXY model on a square lattice, given by the 
Hamiltonian \cite{Villain,TJ},
\begin{equation}
	{\cal 
	H}[\theta({\bf r}_{i})]=\sum_{i,\mu}V\left(\theta({\bf r}_{i}+\hat\mu)-\theta({\bf r}_{i})
	-A_{\mu}({\bf r}_{i}) \right)
	\enspace.
\label{eH}
\end{equation}
Here $\theta({\bf r}_{i})$ is the thermally fluctuating phase 
angle of the planar XY spin on site 
${\bf r}_{i}=x_{i}\hat x+y_{i}\hat y$ ($x_{i}$, $y_{i}$ integers) 
of a $L_{x}\times L_{y}$ periodic square lattice, $\hat\mu=\hat x, \hat y$ label the 
bond directions of the lattice, and $A_{\mu}({\bf r}_{i})$ is the quenched gauge field 
on the bond leaving site ${\bf r}_{i}$ in direction $\hat\mu$ (with 
$A_{-\mu}({\bf r}_{i}+\hat\mu)\equiv -A_{\mu}({\bf r}_{i})$).  For full 
frustration, the $A_{\mu}({\bf r}_{i})$ are constrained so that their 
directed sum 
going counterclockwise around 
any plaquette $P$ of the lattice is fixed (modulus $2\pi$) to,
\begin{equation}
	\sum_{P}A_{\mu}({\bf r}_{i})=\pi\enspace.
\label{ef}
\end{equation}
To implement the constraint of Eq.~(\ref{ef}), we use the specific 
gauge choice,
\begin{equation}
A_{x}({\bf r}_{i})=0,\quad A_{y}({\bf 
r}_{i})=(-1)^{x_{i}}(\pi/2)\enspace.
\label{eA}
\end{equation}

The interaction potential $V(\phi)$ is periodic on $[0, 2\pi)$, 
with a single quadratic minimum at $\phi=0$.  We will take for
$V(\phi)$ the commonly used Villain function \cite{Villain2},
\begin{equation}
	V(\phi)=-T\ln\left[\sum_{m=-\infty}^{\infty}e^{-J(\phi-2\pi 
	m)^{2}/2T}\right]\enspace.
\label{eV}
\end{equation}
The boundary conditions for the phase angles are, in the most
general case, given by, \cite{twist1,twist2}
\begin{equation}
\theta({\bf r}_i+L_{\mu}\hat\mu)-\theta({\bf r}_{i})=\Delta_{\mu}\enspace, 
\label{eBC}
\end{equation}
where  $\Delta_{\mu}\in [0,2\pi)$ is the total twist applied across the system in 
direction $\hat\mu$.  $\Delta_\mu=0$ corresponds to periodic
boundary conditions.  Alternatively, if one makes the change of variables,
$\theta^\prime({\bf r}_i)\equiv \theta({\bf r}_i)-{\bf r}_i\cdot{\bf d}$,
with $d_\mu\equiv\Delta_\mu/L_\mu$, then the system has periodic
boundary conditions in the $\theta^\prime({\bf r}_i)$ and the applied twist
appears as an additive constant to the gauge field, $A_\mu({\bf r}_i)
\to A_\mu({\bf r}_i)+\Delta_\mu/L_\mu$.

\begin{figure}
\epsfxsize=7.5truecm
\epsfbox{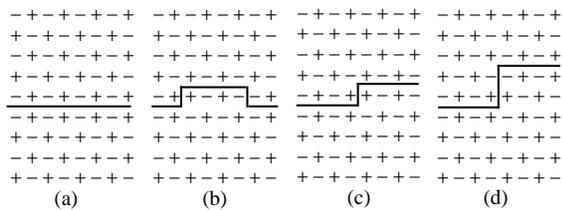}
\caption{Various configurations of the domain wall in a $L\times 
(L+1)$ system: (a) ground state, (b) finite width step of unit height
(kink-antikink pair), (c) isolated kink of unit height, (d) 
isolated kink of height two. A $(+)$ indicates the presence of a vortex in the
XY model, or a charge $q_i=1/2$ in the dual Coulomb gas; a $(-)$ indicates
the absence of a vortex in the XY model, or a charge $q_i=-1/2$ in the dual
CG.  $\hat x$ is the horizontal direction, and $\hat y$
is the vertical direction. 
}
\label{fig1}
\end{figure}

To study the behavior of the Ising-like domain walls we consider systems 
with sizes $L_{x}=L$, $L_{y}=L+1$, with $L$ even.  The odd length $L_y$
forces into the ground state checkerboard pattern of vortices 
a single straight domain wall 
running the length of the system in the $\hat x$ direction.  This is 
illustrated in Fig.~\ref{fig1}a, where a $(+)$ signifies a vortex in the
phase angles $\theta({\bf r}_{i})$, and a $(-)$ signifies the absence of a vortex.

Phase coherence in the FFXY model is most conveniently studied by considering
the dependence of the total free energy $F$ on the total twist 
$\Delta_\mu$ applied across the system (see Eq.~(\ref{eBC})).  In a
phase coherent ordered state, we expect that $F(\Delta_\mu)$ varies with the
twist $\Delta_\mu$; in a phase incoherent disordered state, we expect
$F(\Delta_\mu)$ is independent of $\Delta_\mu$ in the thermodynamic
limit of $L\to\infty$.  The dependence of the free energy on $\Delta_\mu$
is readily obtained by using {\it fluctuating twist boundary conditions} \cite{twist2},
in which one treats the applied twist $\Delta_\mu$ as a thermally fluctuating
degree of freedom.  If $Z$ is the partition function for this ensemble, then the
probability $P(\Delta_\mu)$ of finding a state with a particular twist $\Delta_\mu$ is given by,
\begin{equation}
P(\Delta_\mu)={{\rm e}^{-F(\Delta_\mu)/T}\over Z}\enspace,
\label{ePD}
\end{equation}
and so the free energy with respect to a reference twist $\Delta_{\mu 0}$ is,
\begin{equation}
F(\Delta_\mu)-F(\Delta_{\mu 0}) = -T\ln\left[P(\Delta_\mu)/P(\Delta_{\mu 0})\right]
\enspace.
\label{eFD}
\end{equation}

\begin{figure}
\epsfxsize=7.5truecm
\epsfbox{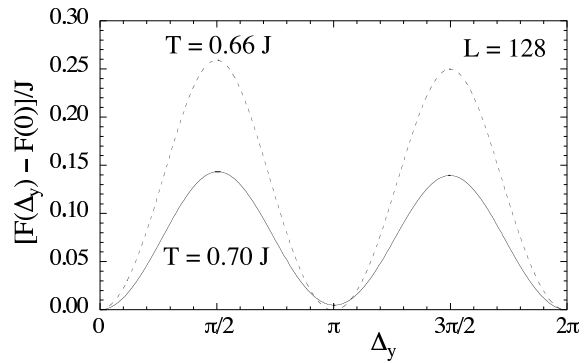}
\caption{Variation of total free energy $F$ with total twist $\Delta_y$ 
applied transverse to the Ising-like domain wall, for two different temperatures
in a system of size $L=128$.
}
\label{fig2}
\end{figure}

The probability $P(\Delta_\mu)$ is directly measured within our fluctuating twist 
Monte Carlo simulation.  We choose the reference twist $\Delta_{\mu 0}$ to be the value of
the twist that minimizes the free energy $F(\Delta_\mu)$.  For the
gauge choice of Eq.~(\ref{eA}), it is straightforward to see to that the minimizing
twist in the $\hat x$ direction is at $\Delta_{x0}=0$.  In our simulations we
keep a fixed twist $\Delta_x=0$, and consider only the dependence of
the free energy on the varying twist $\Delta_y$, transverse to the Ising-like domain
wall that is introduced in our $L\times (L+1)$ systems (see Fig.~\ref{fig1}a).
In Fig.~\ref{fig2} we show sample results from our simulations for
$F(\Delta_y)-F(0)$ vs. $\Delta_y$ at two different values of $T<T_w$, for a system of 
size $L=128$.  We see that $F(\Delta_y)$ has
two equal minima at $\Delta_y=0$ and $\pi$ (i.e. 
periodic and antiperiodic boundary conditions).  
One of these minima corresponds to states where the domain wall sits
at even values of the height $y$, while the other corresponds to states
where the domain wall sits at odd values of the height $y$. 
Below the kink-antikink unbinding transition $T_w$, the system
is in a state of broken translational symmetry; due to the free energy
barrier between the two minima, states in which the
domain wall is at an even height cannot be reached from states in
which the domain wall is at an odd height.
As noted by Korshunov \cite{Korshunov}, this broken symmetry
is restored when phase coherence transverse to the wall
is lost, i.e. when $F(\Delta_y)$ becomes independent of $\Delta_y$, 
and so the free energy barrier between
$\Delta_{y}=0$ and $\Delta_{y}=\pi$ vanishes. 
Alternatively viewed, when the domain wall changes its height by an
odd number, the system acquires an average twist of $\pi$ in the $\hat y$ 
direction.  Thus, restoring the symmetry of domain wall
translations leads to phase angle fluctuations that destroy phase coherence 
transverse to the direction of the wall.

In our numerical work we will use two convenient measures of the
variation of $F(\Delta_\mu)$ with $\Delta_\mu$.  The first is the
helicity modulus\cite{TJ,twist1}, $\Upsilon_\mu$, which measures the curvature
of  $F(\Delta_\mu)$ at its minimum,
\begin{eqnarray}	
	&&\Upsilon_{\mu}(L_{x},L_{y})={L_\mu^2\over L_xL_y}\left.{\partial^{2}F\over\partial\Delta_{\mu}^{2}}
	\right|_{\Delta_{\mu}=0}\nonumber\\
	&=&{1\over L_xL_y}\left\{\sum_{i}\langle V^{\prime\prime}(\phi_{i\mu})\rangle_0
		-{1\over T}\Big\langle\Big[\sum_{i}
	V^{\prime}(\phi_{i\mu})\Big]^{2}\Big\rangle_0\right\}\enspace,
\label{eUps}
\end{eqnarray}
where $\phi_{i\mu}\equiv\phi_{\mu}({\bf r}_i)\equiv \theta({\bf r}_{i}+\hat\mu)-\theta({\bf 
r}_{i})-A_{\mu}({\bf r}_{i})$, $V^{\prime}$ and 
$V^{\prime\prime}$ are the first and second derivatives of the Villain 
function of Eq.~(\ref{eV}), and $\langle\dots\rangle_0$ indicates a thermodynamic
average in the ensemble with fixed twist $\Delta_\mu=0$.  A second measure is,
\begin{equation}
\Delta F = F_{\rm max}-F_{\rm min}=F(\pi/2)-F(0)\enspace,
\label{eDF}
\end{equation}
where $F_{\rm max}$ and $F_{\rm min}$ are the maximum and minimum values
of $F(\Delta_y)$, as $\Delta_y$ is varied at fixed $\Delta_x=0$.  
Since $\Upsilon_\mu$ is an intensive quantity, it should approach a value
independent of system size as $L\to \infty$.  The parameter $\Delta F$
scales as $\Upsilon (\Delta/L)^2 L^D$ in $D$ dimensions, and so for $D=2$
it also becomes independent of system size as $L\to\infty$.


\subsection{The Coulomb gas}

Although our simulations are carried out in the XY variables 
$\theta({\bf r}_i)$, it is helpful to consider the situation from the
viewpoint of the dual CG model of logarithmically interacting
half integer charges \cite{Villain,cg}.  For the case of a fixed total
twist $\Delta_\mu$, the XY Hamiltonian of Eq.~(\ref{eH}) maps onto,
\begin{equation}
{\cal H}_{\rm CG}={\cal H}_0 + {\cal H}_1\enspace.
\label{eHCG}
\end{equation}
${\cal H}_0$ is the logarithmic interaction of the charges,
\begin{equation}
{\cal H}_0={1\over 2}(2\pi J)\sum_{i,j}q_iG({\bf r}_i-{\bf r}_j)q_j\enspace,
\label{eHCG0}
\end{equation}
where $q_i=\pm 1/2$ are the half integer charges, neutrality is imposed, 
$\sum_i q_i=0$, and $G({\bf r})$ is
the 2D periodic lattice Coulomb potential\cite{Villain,cg} with $G({\bf r})\sim -\ln|{\bf r}|$
for large $1\ll |{\bf r}| \ll L/2$.
 ${\cal H}_1$ arises from the fixed twist boundary condition and is given by \cite{Vallat,Bokil},
\begin{equation}
{\cal H}_1=V_x\left(\Delta_x-A_x^0+{2\pi p_y\over L_y}\right)
               + V_y\left(\Delta_y-A_y^0-{2\pi p_x\over L_x}\right)\enspace,
 \label{eHCG1}
 \end{equation}
where $V_x$ and $V_y$ are Villain functions as in Eq.~(\ref{eV}), but with
couplings $J_x=J(L_y/L_x)$ and $J_y=J(L_x/L_y)$ respectively, ${\bf p}$ is the total dipole moment,
\begin{equation}
{\bf p}=\sum_i q_i{\bf r}_i\enspace,
\label{edipole}
\end{equation}
and $A_x^0=\sum_x A_x(x,y=0)$, $A_y^0=\sum_y A_y(x=0,y)$. 
For the gauge choice of Eq.\,(\ref{eA}) we have, 
\begin{equation}
A_x^0=0,\quad A_y^0=(L+1)\pi/2\enspace.
\label{eA0}
\end{equation}

For the ground state as illustrated in Fig.\,\ref{fig1}a,  one has $p_y=0$, and so
again it is easy to see from Eq.\,(\ref{eHCG1}) that the total ground state energy
is minimized when $\Delta_x=0$.  However, for this ground state one has $p_x=L/4$,
hence the ground state energy is minimized when $\Delta_y=(L+1)\pi/2-\pi/2=L\pi/2$.
If the location of the domain wall was  shifted by one unit in height, then the ground
state would have $p_x=-L/4$, and the energy would be minimized when
$\Delta_y=(L+1)\pi/2+\pi/2=L\pi/2+\pi$.  For $L$ even, as we have required,
these two values, modulus $2\pi$, are just equal to $0$ and $\pi$.

The helicity modulus $\Upsilon_y/J$ maps \cite{KT,twist1,twist2} onto the
inverse dielectric function $\epsilon_x^{-1}$ of the CG.  
As noted by 
Korshunov \cite{Korshunov}, in order for the domain wall to move a unit lattice spacing
in height, it must first form a unit step of finite width 
$\ell$; $\ell$ must be even to preserve charge neutrality 
(see Fig.~\ref{fig1}b).
We denote the left hand edge of the step as the {\it kink}, and the 
right hand edge as the {\it antikink}.  As the kink and antikink 
separate out to infinity, the domain wall moves one unit in height. 
As shown by Halsey \cite{Halsey}, a corner in a domain
wall carries with it a net charge of $\pm 1/4$.  The kink, 
consisting of two successive corners with equal $+1/4$ charge, carries a net charge 
of $q_{\rm kink}=+1/2$; the antikink carries a net charge of $q_{\rm kink}=-1/2$.  At low 
temperatures, the logarithmic attraction between the kink and antikink 
charges keeps them bound with a largest separation $\ell_{\mathrm{max}}(T)$.
At higher temperatures, entropy wins out over energy, and there is a
kink-antikink unbinding transition at $T_w$, where 
$\ell_{\mathrm{max}}(T_w)\to\infty$.  Above $T_w$, the 
kink-antikink unbinding leads to diverging dipole fluctuations in the 
$\hat x$ direction, driving $\epsilon_x^{-1}$ (and hence $\Upsilon_y$) 
to zero.

The problem of logarithmically interacting charges in one dimension (1D)
has been treated by Bulgadaev \cite{Bulgadaev}.  Koshunov\cite{Korshunov} 
has applied these results
to the unbinding transition of the kink-antikink pair along
the one dimensional Ising-like domain wall.  To include the screening effect of charge
excitations in the bulk of the system on either side of the Ising-like domain wall,
we take as the coupling between kink-antikink pairs separated at large distance to
be the helicity modulus of the FFXY for an ordinary $L\times L$
system, $\Upsilon(L,L)$, in the limit of large enough $L$.  Applying
Bulgadaev's exact result for the unbinding transition temperature, we conclude,
\begin{equation}
{2\pi \Upsilon(L,L)q_{\rm kink}^2\over T_w}=2, \quad{\rm or}\quad
T_w={\pi\over 4}\Upsilon(L,L)\enspace.
\label{eTw}
\end{equation}

One can reproduce this result using a Kosterlitz-Thouless-like argument \cite{KT}
as follows. In analogy with Lee {\it et al.} \cite{LeeCG}, we consider the total free 
energy to have 
a single ``free'' (i.e. unbound) kink in the domain wall (see 
Fig.~\ref{fig1}c).  Fixing the kink at a given position on the domain 
wall, its free energy (averaging over all other fluctuations) 
is that of an isolated $+1/2$ vortex in a medium with 
phase stiffness $\Upsilon(L,L)$; here we use Bulgadaev's result \cite{Bulgadaev} that 
kink-antikink pairs in 1D do not lead to a renormalization of the kink-antikink
interaction, and so any screening of their interaction is due to charge
excitations in the bulk on either side of the domain wall, and so
accounted for by the large $L$ value of $\Upsilon$.
As $L\to\infty$, the leading 
contribution to this energy is $E=\pi q_{\rm kink}^2\Upsilon\ln L$.
The entropy of the kink is just that associated with its position 
along the domain wall, $S=-\ln L$.  Combining gives 
$F_{\mathrm{kink}}=E-TS=(\pi \Upsilon/4 - T)\ln L$, which as $L\to\infty$ 
gives the instability temperature for the formation of free kinks as
$T_w=\pi \Upsilon/4$, in agreement with Eq.~(\ref{eTw}).

\section{Numerical Results}
  
 \subsection{Helicity modulus}
 
\begin{figure}
\epsfxsize=7.5truecm
\epsfbox{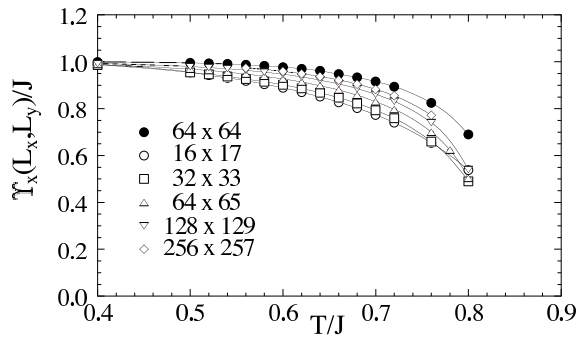}
\caption{Helicity modulus $\Upsilon_{x}$ vs.\ $T$ for different system 
sizes $L_{x}\times L_{y}$.  As $L\to\infty$, $\Upsilon_{x}(L,L+1)$ and 
$\Upsilon_{x}(L,L)$ approach the same curve.
}
\label{fig3}
\end{figure}

\begin{figure}
\epsfxsize=7.5truecm
\epsfbox{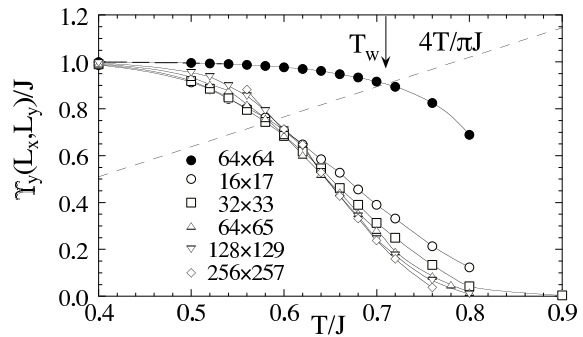}
\caption{Helicity modulus $\Upsilon_{y}$ vs.\ $T$ for different system 
sizes $L_{x}\times L_{y}$.  As $L\to\infty$, $\Upsilon_{y}(L,L+1)$ vanishes at a temperature 
lower than where $\Upsilon_{y}(L,L)$ vanishes.  The intersection of the dashed
line with $\Upsilon_y(L,L)$ indicates the kink-antikink unbinding temperature $T_w\simeq 0.71J$
predicted by Eq.~(\protect\ref{eTw}).
}
\label{fig4}
\end{figure}

We now present our numerical results.
At each temperature, our simulations consist of typically $10^8 - 10^9$ ordinary MC
passes through the entire lattice for the largest system sizes.  In Figs.~\ref{fig3} and 
\ref{fig4} we plot the 
helicity moduli $\Upsilon_x$ and $\Upsilon_y$ vs.\ $T$, as computed by Eq.~(\ref{eUps}) 
in an ensemble with fixed twists $\Delta_x=\Delta_y=0$.
We show an ``ordinary'' case (no domain wall at $T=0$) of size 
$64\times 64$, which is large enough that any finite size effects are
negligible for the temperatures shown.  In comparison, we also show 
several ``anomalous'' cases (percolating domain wall at $T=0$) of sizes $L\times (L+1)$.  
For the ordinary case, $\Upsilon_x=\Upsilon_y$ and 
the bulk transition (where the $\Upsilon_\mu$ jump discontinuously to zero
in the thermodynamic limit) is \cite{Olsson2} at $T_{KT}\simeq 0.81J$.
In comparison, as $L$ increases in the anomalous case, $\Upsilon_x$ 
(parallel to the domain wall) in Fig.\,\ref{fig3} approaches the value of the ordinary 
case, and so presumably vanishes at the same
$T_{KT}$.  However the curves of $\Upsilon_y$ (transverse to the 
domain wall) in Fig.\,\ref{fig4} clearly decrease below that of the ordinary case, and 
presumably vanish in the thermodynamic limit at a lower $T_w$.  


The reduction seen in $\Upsilon_y$ for the $L\times (L+1)$ systems, as compared
to the $L\times L$ system, shown in Fig.\,\ref{fig4}, is due to the
kinks at the Ising-like domain wall.  To explicitly see this, we can consider 
the helicity modulus at {\it finite wavevector}, $\Upsilon_y(k_x)$, defined as the response
to a small sinusoidal perturbation in the vector potential $A_y({\bf r}_i)$.  If
we take,
\begin{equation}
A_y({\bf r}_i)\to A_y({\bf r}_i)+\sum_{k_x}\delta A_{k_x} {\rm e}^{ik_x x_i}\enspace
\label{edA}
\end{equation}
then $\Upsilon_y(k_x)$ is defined by \cite{Olsson2,Upsk},
\begin{equation}
\Upsilon_y(k_x)={1\over L_xL_y}\left.{\partial^2 F\over\partial\delta A_{k_x}
\partial\delta A_{-k_x}}\right|_{\delta A_{k_x}=0}\enspace.
\label{eUpsk}
\end{equation}
In view of the discussion following Eq.~(\ref{eBC}), equating the application of
a uniform twist $\Delta_\mu$ to the addition of a constant to the 
gauge field $A_\mu$, the helicity modulus $\Upsilon_y$ of Eq.~(\ref{eUps}) can also be viewed 
as the {\it zero wavevector} helicity $\Upsilon_y(k_x=0)$.
In the CG representation, $\Upsilon_y(k_x)$ becomes the usual
formula for the wavevector dependent inverse dielectric function \cite{twist1,twist2},
\begin{equation}
\Upsilon_y(k_x)/J = 1-{4\pi^2 J\over T}{\langle q({k_x})q({-k_x})\rangle\over L_xL_y\>k_x^2}
\enspace,
\label{eUpskCG}
\end{equation}
where $q({\bf k})=\sum_i {\rm e}^{i{\bf k}\cdot{\bf r}_i}q_i$ is the Fourier transform
of the charge distribution.

Unlike $\Upsilon_y$ of Eq.~(\ref{eUps}), which measures the response to a uniform twist
applied at the boundaries, $\Upsilon_y(k_x)$ measures the response to a spatially varying
twist applied throughout the bulk of the system.  For a homogeneous system with
{\it periodic boundary
conditions}, one in general expects $\Upsilon_y = \lim_{k_x\to 0}\Upsilon_y(k_x)$, since 
the spatially varying twist becomes uniform as $k_x\to 0$, and $\delta A_{k_x}\to\Delta_y/L_y$.
For {\it free boundary conditions} however, where the phase angle $\theta(x,L_y)$ is not
coupled to the phase angle $\theta(x,0)$, this equality does not hold.  For free boundary
conditions, the absence of any constraint (such as in Eq.~(\ref{eBC}))
relating $\theta(x,L_y)$ to $\theta(x,0)$
means that the phase angles are free to untwist any additive constant to the
gauge field, $A_y({\bf r}_i)\to A_y({\bf r}_i)+\Delta_y/L_y$,
by choosing $\theta(x,y+1)-\theta(x,y)=\Delta_y/L_y$; hence if one computes
$\Upsilon_y$ by Eq.~(\ref{eUps}) in a free boundary ensemble, 
one necessarily has $\Upsilon_y=0$ at any temperature.
For the spatially varying twist of Eq.~(\ref{edA})
however, no such transformation is possible since the perturbing twist is a strictly
transverse vector function, while the phase angle differences give a strictly longitudinal
vector function. In this case one finds that $\lim_{k_x\to 0} \Upsilon_y(k_x)$ has the
same value, as $L\to\infty$, that one has for the system with periodic boundary
conditions.

We expect a similar effect to be true in our present case.  The kink-antikink pairs confined
to the one dimensional  Ising-like domain wall can be viewed as a relaxation of the
boundary condition.  They can unwind, or soften the energy of  a uniform 
twist $\Delta_y$ applied at the boundary, but cannot unwind a spatially varying 
twist $\delta A_{k_x}$ applied throughout the bulk of the system.  We therefore expect
that, as $L\to\infty$, $\lim_{k_x\to 0} \Upsilon_y(k_x)$ will equal the value of 
$\Upsilon_y$ obtained for an ordinary $L\times L$ system, representing the stiffness
of the bulk of the system on either side of the domain wall; $\Upsilon_y$, however,
will be a lower value including effects due to the polarization of the kink-antikink
pairs localized to the domain wall.

\begin{figure}
\epsfxsize=7.5truecm
\epsfbox{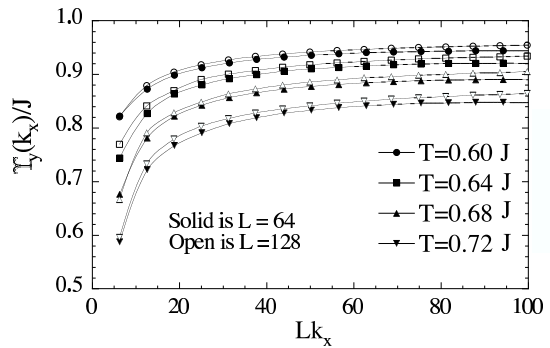}
\caption{Wave vector dependent helicity modulus $\Upsilon_{y}(k_x)$ vs.\ $Lk_x$ 
at several different temperatures $T$ for system sizes $L\times (L+1)$, with $L=64$ (solid
symbols), and $L=128$ (open symbols).
The data for the different $L$ collapse to a common curve at each $T$.
}
\label{fig5}
\end{figure}

\begin{figure}
\epsfxsize=7.5truecm
\epsfbox{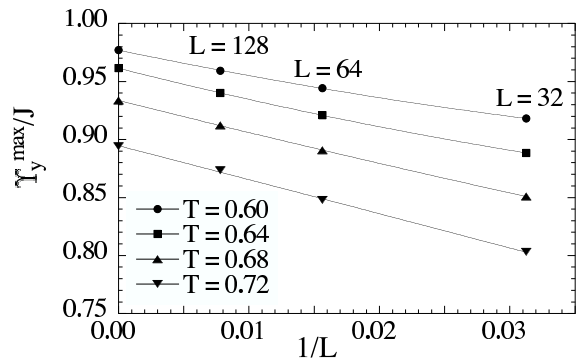}
\caption{$\Upsilon_y^{\rm max}\equiv \max_{k_x}[\Upsilon_y(k_x)]$ vs. $1/L$
at several different temperatures $T$.  The values at $1/L=0$ correspond to 
$\Upsilon_y$ for an ordinary $64\times 64$ system.  Solid lines are fits to a
quadratic polynomial.
}
\label{fig6}
\end{figure}

In Fig.~\ref{fig5} we plot, at several different temperatures around $T_w$, 
$\Upsilon_y(k_x)$ for finite $k_x$ for $L\times (L+1)$ systems with sizes $L=64$ and $128$.
When we plot the results versus the scaled axis of $Lk_x$, we see that the data for
the two different system sizes collapse to essentially a common curve, $u(Lk_x)$, at each temperature.  As $L\to\infty$, such scaling implies that
$\lim_{k_x\to 0}\left[\lim_{L\to\infty}\Upsilon_y(k_x)\right]
=\lim_{\kappa\to\infty}u(\kappa)$, is different from $\Upsilon_y$, just as we
have argued.  We may estimate $\lim_{\kappa\to\infty} u(\kappa)$ by
taking the maximum value of $\Upsilon_y(k_x)$ for each system size $L$.
In Fig.~\ref{fig6} we plot $\Upsilon_y^{\rm max}\equiv \max_{k_x}[\Upsilon_y(k_x)]$
versus $1/L$ for the same temperatures as shown in Fig.~\ref{fig5}.  We give results
for $L=32,64$ and $128$.  At $1/L=0$ we plot the value of $\Upsilon_y(L,L)$
for $L=64$, representing the large $L$ limit for the helicity modulus of the
ordinary system without the Ising domain wall.  We see that the values of 
$\Upsilon_y^{\rm max}$ extrapolate perfectly to $\Upsilon_y(L,L)$ as $1/L\to 0$.  
This confirms the following conclusion.  As $L\to\infty$ in an $L\times (L+1)$ system, 
the finite wavevector helicity $\Upsilon_y(k_x)$
is equal to the corresponding helicity modulus of an ordinary $L\times L$ system for {\it all}
finite values of $k_x$; this measures the stiffness of the bulk of the system on
either side of the Ising domain wall, and is unaffected by the kinks on the wall.
However the zero wavevector response $\Upsilon_y$ to a uniform twist $\Delta_y$, 
shown in Fig.~\ref{fig4},
is softened, and above $T_w$ reduced to zero, by the polarization of the kinks
on the domain wall.


\subsection{Finite size dependence}

Returning to Fig.~\ref{fig4}, we have indicated the kink-antikink unbinding
transition temperature $T_w$, as predicted in Eq.~(\ref{eTw}), by the intersection
of the line $4T/\pi J$ with the helicity modulus of the ordinary $L\times L$ system.
This gives an estimate of $T_w\simeq 0.71 J$.  This value occurs noticeably above
the point where many of the curves $\Upsilon_y(L,L+1)$ appear to cross.
Such a crossing point, if remaining constant as $L$ increases, is generally taken
as an estimate for the phase transition in a 2D XY system.  To explain the
difference between this crossing point at $\sim 0.60 J$ and the above estimate
$T_w\simeq 0.71 J$, we need to examine the finite size dependence of $\Upsilon_y(L,L+1)$
more carefully.  

\begin{figure}
\epsfxsize=7.5truecm
\epsfbox{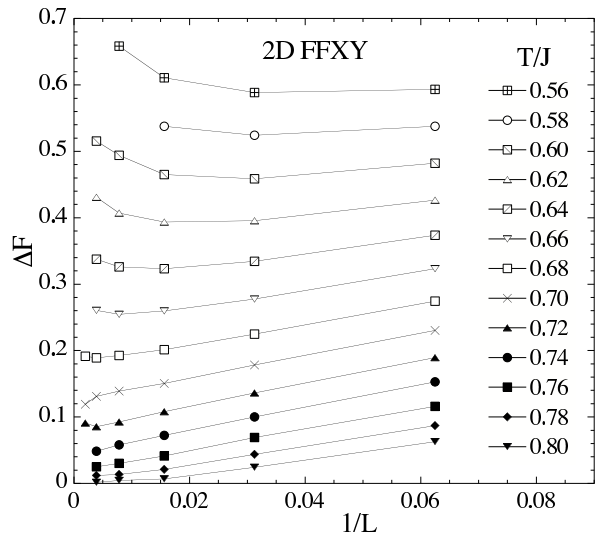}
\caption{$\Delta F$ of Eq.~(\ref{eDF}) vs. $1/L$ at various temperatures
for the 2D FFXY model of size $L\times (L+1)$.
}
\label{fig7}
\end{figure}

To get the most accurate results, we have found it better to work in the
fluctuating twist ensemble and compute the phase coherence parameter
$\Delta F$ of Eq.~(\ref{eDF}), rather than work with periodic boundary
conditions and measure $\Upsilon_y$.  In Fig.~\ref{fig7} we plot our
results for $\Delta F$ versus system size $1/L$, for various temperatures $T$
above and below $T_w$.  We use $L\times (L+1)$ systems with $L$ ranging
from $32$ up to $512$.  We note that at low $T$, the behavior of $\Delta F$
is {\it non-monotonic}; as $L$ increases, the system first softens with
a decreasing $\Delta F$, but then stiffens again as $\Delta F$ reaches a
minimum and then increases.  We denote the system size at this minimum
by $\xi_k$.  Since the system becomes stiffer on length scales $L>\xi_k$,
we assume that $\xi_k$ determines the size of the largest kink-antikink pairs
on the domain wall.  At higher $T$, the minimum in $\Delta F$ disappears,
and $\Delta F$ continues to decrease as $L$ increases.  The temperature that
separates the two behaviors is somewhere between $0.68 J$ and $0.74 J$,
in good agreement with the result $T_w\simeq 0.71 J$ from Fig.~\ref{fig4},
based on the theoretical prediction of Eq.~(\ref{eTw}).

As a further check on our results, we have also directly simulated a 1D neutral
system of logarithmically interacting charges $q_{\rm kink}=\pm 1/2$.
The 1D interaction
potential is taken  as the 2D $L\times L$ lattice Coulomb potential, as in
Eq.(\ref{eHCG0}), but with height separation fixed at $y=0$.  
We use an interaction coupling constant of $2\pi\Upsilon_y(L,L)$, 
with $\Upsilon_y(L,L)$ obtained
from our simulations of the ordinary $L\times L$ 2D FFXY for $L=64$,
in order to model as closely as possible the interaction between kinks on
the domain wall in the true $L\times (L+1)$ FFXY system.  
Within this 1D simulation we measure the normalized histogram of the total 
dipole moment, $P(p_x)$,
and use it to construct \cite{Gupta} what would be the free energy $F(\Delta_y)$
of the corresponding $L\times (L+1)$ 2D FFXY system,
\begin{equation}
F(\Delta_y)=-T\ln\left[\sum_{p_x}P(p_x){\rm e}^{-V_y\left(\Delta_y-A_y^0-
{2\pi p_x\over L}\right)/T}\right]+{\rm const.}\enspace,
\label{eFD1D}
\end{equation}
where $V_y$ is the Villain function as in Eq.~(\ref{eHCG1}), and ``const."
is a constant term independent of $\Delta_y$.

\begin{figure}
\epsfxsize=7.5truecm
\epsfbox{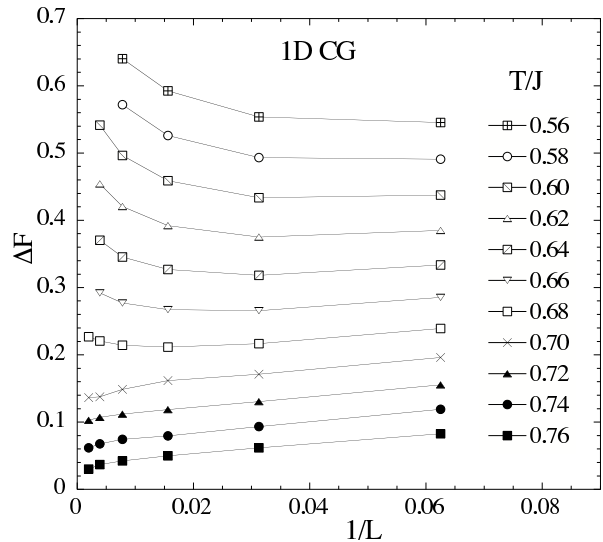}
\caption{$\Delta F$ of Eq.~(\ref{eDF}) vs. $1/L$ at various temperatures
for as estimated from the dipole histogram of a 1D interacting kink model (see text).
}
\label{fig8}
\end{figure}

In Fig.~\ref{fig8} we plot the resulting $\Delta F=F(\pi/2)-F(0)$ versus 
$1/L$ for the same temperatures and sizes $L$ as in Fig.~\ref{fig7}.
The agreement between Figs.~\ref{fig7} and \ref{fig8} is not exact, since
the coupling between kinks is only equal to $2\pi\Upsilon_y(L,L)$ on {\it large}
length scales; the true screening of the kink interaction due to charge
excitations in the bulk on either side of the domain wall is length scale dependent.
Moreover, the domain wall in the 2D FFXY model is not a strictly straight one
dimensional line; the roughness of the domain wall (see following section)
means that height fluctuations can add to the distance of separation 
between kinks.
Nevertheless, the agreement is qualitatively very good, indicating again  that
it is the polarization of kink-antikink pairs along the Ising domain wall that
is responsible for the decrease in the phase stiffness transverse to the domain wall.


\begin{figure}
\epsfxsize=7.5truecm
\epsfbox{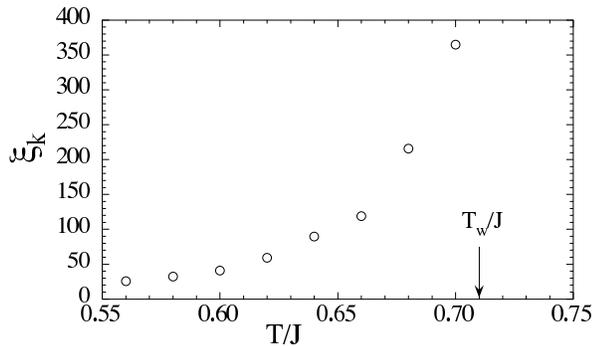}
\caption{Kink-antikink correlation length $\xi_k$ vs. $T$, obtained from
the data of Fig.~\ref{fig7}.
}
\label{fig9}
\end{figure}

One evident feature of both Figs.~\ref{fig7} and \ref{fig8} is the very large
finite size effect.  The asymptotic behavior of $\Delta F$ only sets in
at quite large length scales.  Equivalently, the correlation length of the
kink-antikink pairs, $\xi_k$, grows large well below $T_w\simeq 0.71 J$.  
Fitting the data of Fig.~\ref{fig7} to a quadratic in $\ln L$, and determining
$\xi_k$ from the minima of these fitted curves, we plot the resulting $\xi_k$
versus $T$ in Fig.~\ref{fig9}.
The rapidly growing $\xi_k$ means that
it is difficult to get a very precise estimate of $T_w$ directly from the data 
of Fig.~\ref{fig7}, without going to prohibitively large system sizes $L\gg512$.
We believe that this is also the reason that Lee {\it al}. \cite{LeeCG}
report a lower value\cite{foot3} of $T/J = 2\pi(0.09\pm 0.01) = 0.57\pm 0.06$ 
for the ``roughening transition" temperature in their dual 2D
Coulomb gas.  Lee {\it et al}.'s simulations are on a lattice of size $L=64$.
From Fig.~\ref{fig9} we see that $\xi_k\sim 32$ at $T/J\sim 0.58$, hence
kinks already look unbound above this temperature for such a small system size.
The large values of $\xi_k$ can also be compared to other length scales in 
the system.  The correlation length of the Ising-like order parameter, $\xi_I$,
gives the typical size of an Ising-like domain excitation in an ordinary $L\times L$
FFXY model.  From Ref.~\onlinecite{Olsson2} (see Fig.~15) we find that $\xi_I$
is quite small below $T_w\simeq 0.71 J$; in particular $\xi_I<2.5$ for $T<0.77 J$.
Thus we expect that kink-antikink pairs on the domain wall enclosing a typical
thermally excited Ising-like domain are always effectively unbound.  The
energetics of kink-antikink unbinding will only effect the 
morphology of domains that are {\it much} larger than those due to typical
thermal excitations.

\begin{figure}
\epsfxsize=7.5truecm
\epsfbox{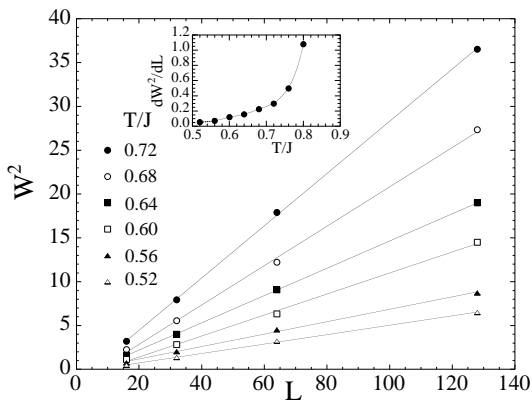}
\caption{Domain wall width squared, $W^2$, vs.\ $L$ for several
different $T$.  Straight lines are linear fits, showing rough
domain walls at all $T$, even below $T_w\simeq 0.71 J$.  Inset shows the slopes of 
the curves, $dW^2/dL$, vs.\ $T$.
}
\label{fig10}
\end{figure}

\subsection{Roughening}

Finally, we consider the roughness of the domain wall.  As noted by
Koshunov \cite{Korshunov}, an isolated step on a domain wall of height two (see 
Fig.~\ref{fig1}d) carries no net charge, 
each corner giving opposite $1/4$ charges that therefore cancel.
Such isolated height of two steps therefore cost finite energy, and 
should roughen the domain wall at any finite temperature.  To verify 
this we have explicitly measured the domain wall width squared,
\begin{equation}
	W^2={1\over N_s}\sum_s\left\langle(y(s)-\bar{y})^2\right\rangle \enspace,
\label{eW}
\end{equation}
where $s$ is an index that counts horizontal length traveled along the 
domain wall ($s=x$ if there are no overhangs), $y(s)$ is the height of 
the domain wall at index $s$, $\bar{y}$ is  the average height of the 
domain wall in the particular configuration, and $N_s$ is the number of
horizontal links in the domain wall ($N_s=L$ if there are no overhangs).  
To avoid problems with periodic 
boundary conditions, we always measure the domain wall height as
relative to some initial starting position.  In Fig.~\ref{fig10} we plot
$W^2$ vs.\ $L$ for various temperatures.  The  
linear growth in $W^2$ as $L$ increases indicates that the domain
wall is rough for all temperatures shown, including $T<T_w$. 
The domain wall diffusion constant, $dW^2/dL$, shows no observable
singularity at $T_w\simeq 0.71 J$ (see inset to Fig.~\ref{fig10}).


\section{Discussion}

Our numerical results demonstrate the existence of the unbinding transition
for kink-antikink pairs along the domain wall of Ising-like excitations
in the 2D FFXY model, and that the behavior of this transition is in
good agreement with the theoretical predictions of Korshunov\cite{Korshunov}.  We show
that domain walls are rough at all temperatures, and therefore argue that
the ``roughening transition" claimed by Lee and co-workers\cite{LeeXY,LeeCG} is really the
kink-antikink unbinding transition.
We show that the effects of kink-antikink pairs are not readily apparent
for Ising-like domains such as are typically present due to thermal excitation;
because of the large length $\xi_k$, such effects are important only for
much larger domains.

One of Korshunov's main motivations for investigating the kink-antikink
unbinding transition was to argue that such a transition necessarily implies  
the existence of two separate bulk transitions, $T_{KT}<T_{I}$.  
We now present our own thoughts on this issue.  
In the original paper by Teitel and Jayaprakash \cite{TJ}, two possibilities 
were considered, $T_{KT}<T_I$ and $T_{KT}=T_I$.  In discussing the 
first case, Teitel and Jayaprakash presented the following scenario.  
In terms of the dual CG 
model, the helicity modulus $\Upsilon$ gets reduced from its $T=0$ value
by fluctuations that produce dipole moments.  If Ising 
domains of typical size $\xi_I$ carried a total dipole moment
proportional to their size, they would drive $\Upsilon\to 0$  
{\it continuously} due to the diverging $\xi_I$ as $T\to T_I$
from below.  However, in addition to these domain 
excitations, there are also pair excitations.
The  original Kosterlitz-Thouless instability criteria would imply 
that pairs unbind once $\Upsilon$ falls below the critical value 
$\Upsilon(T)=2T/\pi$, which must happen at some $T_{KT}$ {\it below} 
$T_I$.  However, if one computes domain energies at $T=0$ (and 
presumably the same holds for domain free energies at low $T$),
one finds that it is only the domains with {\it vanishing} total 
dipole moment that have energies which scale with the perimeter; i.e. 
the only domains which are ``Ising-like'' are those which carry no dipole moment 
and so cannot give any reduction in $\Upsilon$!  Fortunately, once 
$T$ increases above $T_{w}$, kink-antikink pairs on the boundary of 
the domain are free to unbind, and the Ising domains can now acquire 
large dipole moments at no cost in free energy.  The scenario of 
Teitel and Jayaprakash is now restored.  This conclusion is in complete
agreement with the numerical work of Olsson \cite{Olsson1,Olsson2},
who finds two distinct transitions at $T_{KT}<T_I$, and argues that the
non Ising-like critical behavior claimed by some \cite{single} at $T_I$ is in fact
an artifact of finite size effects.

The kink-antikink unbinding transition may also have implications for the 
non-equilibrium steady state behavior of the system when the vortices 
are driven by a uniform force, such as is the case for a fully frustrated
Josephson junction array in an applied uniform d.c. current $I$.
Since $I$ couples linearly to the total
dipole moment of a domain $p$, the force will can can lead to an instability \cite{MT}
in domain growth provided the free energy of exciting the 
domain scales less than linearly with $p$.  
For $T<T_w$, the binding of kinks to antikinks prevents large
dipole moments from building up on domains.  The Ising-like domains, whose
free energy scales like the domain length $\ell$, are only those
domains whose total dipole moment vanishes, and hence these
domains remain stable, and the Ising-like order should persist at
small drives.  For $T>T_w$, kink-antikink pairs can unbind, 
resulting in domains whose dipole moment may scale at least
proportional to their length $\ell$.  In this case, when the current
exceeds an amount proportional to the Ising domain surface tension,
domains will become unstable to growth and the Ising-like order
will be destroyed.  

\section*{Acknowledgements}

We wish to thank S. E. Korshunov for his very helpful correspondence 
and comments on an earlier version of this manuscript.
This work was supported by the
Engineering Research Program of the Office of Basic Energy Sciences
at the Department of Energy grant DE-FG02-89ER14017 and the
Swedish Research Council Contract No. 2002-3975.  
Travel between Rochester and Ume{\aa} was supported by 
grants NSF INT-9901379 and STINT 99/976(00).


\begin{thebibliography}{99}
\bibitem{foot3} We have converted temperature scales between the CG 
model of Ref.~\protect\onlinecite{LeeCG} and our FFXY model, 
$T_{XY}=2\pi T_{CG}$.
\end{thebibliography}
\end{document}